\documentclass[11pt]{article}
\usepackage{graphicx}
\usepackage{cite}
\usepackage{amssymb}
\usepackage{epsfig}
\usepackage{epstopdf}
\usepackage{hep}
\usepackage{feynmp}
\usepackage{multicol}
\usepackage{appendix}
\usepackage{subfigure}
\usepackage{color}
\usepackage{ulem}
\usepackage{float}
\makeatletter
\def\slash#1{{\mathpalette\c@ncel{#1}}} 
\makeatother  

\newcommand{\bfc}{\begin{figure}\begin{center}}
\newcommand{\efc}{\end{center}\end{figure}}

\newcommand{\fig}[2]{\scalebox{#1}{\includegraphics{#2}}}

\newcommand\beq{\begin{eqnarray}}
\newcommand\eeq{\end{eqnarray}}

\begin{document}

\title{{\bf Transverse single-spin asymmetries in proton-proton collisions at the AFTER@LHC experiment}}

\author{K.~Kanazawa$^{1}$, Y.~Koike$^{2}$, A.~Metz$^{1}$, and D.~Pitonyak$^{3}$
 \\[0.3cm]
{\normalsize\it $^1$Department of Physics, SERC,
  Temple University, Philadelphia, PA 19122, USA} \\[0.15cm]
{\normalsize\it $^2$Department of Physics, Niigata University,
Ikarashi, Niigata 950-2181, Japan} \\[0.15cm]
{\normalsize\it $^3$RIKEN BNL Research Center,
                 Brookhaven National Laboratory,
                 Upton, New York 11973, USA} 
}

\date{\today}
\maketitle

\begin{abstract}
\noindent
We present results for transverse single-spin asymmetries in proton-proton collisions at kinematics relevant for AFTER, a proposed fixed-target experiment at the Large Hadron Collider.  These include predictions for pion, jet, and direct photon production from analytical formulas already available in the literature.  We also discuss specific measurements that will benefit from the higher luminosity of AFTER, which could help resolve an almost 40-year puzzle of what causes transverse single-spin asymmetries in proton-proton collisions.
\end{abstract}

%
%
\section{Introduction} \label{s:intro}

Transverse single-spin asymmetries (TSSAs), denoted $A_N$, have been a fundamental observable since the mid-1970s to test perturbative Quantum Chromodynamics (pQCD). Such measurements were first conducted at FermiLab, where large effects were found in $p\hspace{0.025cm}Be\to\Lambda^\uparrow X$~\cite{Bunce:1976yb}.  These results contradicted the na\"{i}ve collinear parton model, which said that $A_N$ should be extremely small~\cite{Kane:1978nd}, and doubts were raised as to whether pQCD can actually describe these reactions~\cite{Kane:1978nd}.  However, in the 1980s it was shown that if one went beyond the parton model and included collinear twist-3 (CT3) quark-gluon-quark correlations in the nucleon, substantial TSSAs could be generated~\cite{Efremov:1981sh}.  In the 1990s this CT3 approach was worked out in more detail for proton-proton collisions, first for direct photon production~\cite{Qiu:1991pp,Qiu:1991wg,Korotkiian:1995vf} and then for pion production \cite{Qiu:1998ia}.  Over the last decade, several other analyses furthered the development of this formalism --- see \cite{Eguchi:2006qz,Eguchi:2006mc,Kouvaris:2006zy,Koike:2006qv,Koike:2007rq,Koike:2009ge,Kanazawa:2010au,Kanazawa:2013uia,Metz:2012ct,Beppu:2013uda} and references therein. During the same time, another mechanism was also put forth to explain TSSAs in proton-proton collisions.  This approach involves the Sivers~\cite{Sivers:1989cc}, Collins~\cite{Collins:1992kk}, and Boer-Mulders~\cite{Boer:1997nt} transverse momentum dependent (TMD) functions and became known as the Generalized Parton Model (GPM) --- see~\cite{Anselmino:1994tv,Anselmino:1998yz,Anselmino:2005sh,Anselmino:2012rq,Anselmino:2013rya} and references therein. (We mention that, since most likely a rigorous factorization formula involving TMD functions does not hold for single-inclusive processes (which have only one scale), the GPM can only be considered a phenomenological model.) In addition to all of this theoretical work, many experimental measurements of $A_N$ have been performed at proton-(anti)proton accelerators~\cite{Adams:1991rw,Krueger:1998hz,Adams:2003fx,Adler:2005in,Lee:2007zzh,:2008mi,Adamczyk:2012qj,Adamczyk:2012xd,Bland:2013pkt,Adare:2013ekj}.  Most of the experimental data in the more negative $x_F$ region\footnote{Throughout the paper we will use the convention $x_F = 2\,l_{z}/\sqrt{S}$, where $l$ is the momentum of the outgoing particle, and the transversely polarized proton moves along the $-z$-axis.  That is, $x_F\to -1$ means large momentum fractions $x^\uparrow$ of the parton probed inside the transversely polarized proton.  This setup causes $x_F$ to be opposite in sign to the one used in collider experiments (like those at RHIC).} has come in the form of light-hadron asymmetries $A_N^h$, e.g., $h=\pi$, $K$, $\eta$, with the exception of the jet asymmetry $A_N^{jet}$ measured a few years ago at the Relativistic Heavy Ion Collider (RHIC) by the A$_N$DY Collaboration~\cite{Bland:2013pkt}.  Plans are also in place to measure the direct photon asymmetry $A_N^\gamma$ at RHIC by both the PHENIX Collaboration and the STAR Collaboration \cite{PHENIX:BeamUse,STAR:BeamUse,Aschenauer:2015eha}.

Although much progress has been made in understanding TSSAs, there is not a definitive answer on what their origin might be.  In the CT3 approach it was assumed for many years that a soft-gluon pole (SGP) chiral-even quark-gluon-quark ($qgq$) matrix element called the Qiu-Sterman (QS) function $T_F(x,x)$ was the main cause of $A_N^\pi$~\cite{Qiu:1998ia,Kouvaris:2006zy}.  However, this led to a so-called ``sign mismatch'' between the QS function and the TMD Sivers function $f_{1T}^\perp$ extracted from semi-inclusive deep-inelastic scattering (SIDIS)~\cite{Kang:2011hk}.  This issue could not be resolved through more flexible parameterizations of the Sivers function~\cite{Kang:2012xf}.  Moreover, the authors of Ref.~\cite{Metz:2012ui} argued, by looking at $A_N$ data on the target TSSA in inclusive DIS~\cite{Airapetian:2009ab,Katich:2013atq}, that $T_F(x,x)$ cannot be the main source of $A_N^\pi$.  This observation led us last year in Ref.~\cite{Kanazawa:2014dca} to analyze $A_N^\pi$ by including not only the QS function but also the fragmentation mechanism, whose analytical formula was first fully derived in~\cite{Metz:2012ct}\footnote{The so-called ``derivative term'' was first computed in~\cite{Kang:2010zzb}.} (see also~\cite{Kanazawa:2013uia,Gamberg:2014eia,Kanazawa:2014tda} for fragmentation terms in other processes).  We found in this situation for the first time in pQCD that one can fit all RHIC high transverse momentum pion data very well without any sign-mismatch issue.  Furthermore, we showed that a simultaneous description of TSSAs in $p^\uparrow p\to\pi X$, SIDIS, and $e^+e^-\to h_1h_2 X$ is possible.  Nevertheless, more work must be done to confirm/refute this explanation and its predictions.  We mention that in the GPM, one cannot draw a definitive conclusion as to whether the Sivers or Collins mechanism is the main cause of $A_N^\pi$~\cite{Anselmino:2012rq,Anselmino:2013rya}.\footnote{In principle the Boer-Mulders function and gluon Sivers function can also contribute in the GPM formalism, but these pieces have not been analyzed in the literature.}  This is due to the theoretical error bands being too large, since the associated TMD functions are mostly unconstrained in the large-$x^\uparrow$ regime covered by the data~\cite{Anselmino:2012rq,Anselmino:2013rya}.  For a detailed discussion of the GPM formalism and its predictions for the AFTER experiment, see Ref.~\cite{Anselmino:2015eoa}.

In addition, in order to have a complete knowledge of TSSAs, it is important to have a ``clean'' extraction of the QS function from observables like $A_N^{jet}$ and $A_N^\gamma$ that do not have any fragmentation contributions.\footnote{We will ignore photons coming from fragmentation~\cite{Gamberg:2012iq}, which can be largely suppressed by using isolation cuts.}  (For recent analyses of $A_N^\gamma$ in $p^\uparrow A$ collisions, see Refs.~\cite{Kovchegov:2012ga,Schafer:2014zea}.) This is necessary in order to help resolve the sign-mismatch issue and better understand the role of re-scattering effects in the nucleon.  The jet asymmetry has been studied in~\cite{Kouvaris:2006zy,Kang:2011hk,Kanazawa:2012kt,Gamberg:2013kla} and the direct photon asymmetry has been investigated in~\cite{Qiu:1991wg,Kouvaris:2006zy,Ji:2006vf,Koike:2006qv,Koike:2011nx,Kanazawa:2011er,Gamberg:2012iq,Kanazawa:2012kt,Gamberg:2013kla,Kanazawa:2014nea}.  It is important to point out that other contributions to $A_N^{jet}$ and $A_N^\gamma$ exist besides the one from the (SGP $qgq$ chiral-even) QS function.  These other pieces include (i) soft-fermion pole (SFP) chiral-even $qgq$ functions, (ii) SGP and SFP $qgq$ chiral-odd functions, and (iii) SGP tri-gluon functions. For $A_N^\gamma$ the numerical analyses in~\cite{Kanazawa:2012kt,Kanazawa:2014nea} show (i) is negligible for $x_F<0$ while the study in~\cite{Kanazawa:2014nea} draws a similar conclusion for (ii) as does the work in~\cite{Koike:2011nx} for (iii).  That is, for $A_N^\gamma$ the QS function dominates the asymmetry.  We mention that at present in the GPM, $A_N^\gamma$ is predicted to have the {\it opposite sign} to that from the CT3 approach~\cite{Anselmino:2013rya}.  Therefore, as was emphasized in~\cite{Kanazawa:2014nea}, this observable could allow us for the first time to clearly distinguish between the two frameworks as well as learn about the process dependence of the Sivers function~\cite{Collins:2002kn}, which is a feature of this non-perturbative object that is crucial to our current understanding of TMD functions.

For $A_N^{jet}$ the conclusions as to which piece dominates are not as clear.  The study in~\cite{Kanazawa:2000hz} provides evidence that (ii) should be 
small in the whole $x_F$-region.  The work in~\cite{Kanazawa:2012kt} shows the same is most likely true for (i), but that analysis suffers from the sign-mismatch issue.  Also, in~\cite{Beppu:2013uda} there is an indication that (iii) could be significant.  Therefore, it will be necessary to re-assess the impact of (i) and (iii) on $A_N^{jet}$.  Nevertheless, one can gain insight into these other terms by looking at the contribution from the QS function and comparing it with data.

Given the open issues that still remain, it is an opportune time for the Large Hadron Collider (LHC) to produce data on TSSAs in proton-proton collisions via the AFTER experiment.  These measurements will not only add to the data from FermiLab, AGS, and RHIC, but also, through the high luminosity of the experiment~\cite{Brodsky:2012vg,Lansberg:2014myg}, probe certain features that remain ambiguous.  For example, the behavior of $A_N^\pi$ at large pion transverse momentum $l_T$ appears to fall off very slowly (or is even flat), a feature which theory says should persist to high $l_T$~\cite{Kanazawa:2012kt,Kanazawa:2014dca,Anselmino:2013rya,Anselmino:2015eoa} (see also~\cite{Anselmino:2000vs} in the context of $\Lambda^\uparrow$ production).  However, the data from RHIC~\cite{Heppelmann:2013ewa} has too large error bars (or not enough statistics) in this high-$l_T$ region to ascertain whether or not this is true.  Also, $A_N^{jet}$ measured by A$_N$DY~\cite{Bland:2013pkt} has large error bars as $x_F$ becomes more negative, which makes it difficult to determine whether or not the QS function alone can describe that data.  Moreover, as previously mentioned, $A_N^\gamma$ has never been measured before, yet it could be a tremendous opportunity to learn about the process dependence of the Sivers function and distinguish between the CT3 and GPM frameworks.  Already, PHENIX and STAR plan to carry out such experiments~\cite{PHENIX:BeamUse,STAR:BeamUse,Aschenauer:2015eha}.

Therefore, in this paper we give predictions within the CT3 formalism for $A_N^\pi$, $A_N^{jet}$, and $A_N^\gamma$ at AFTER@LHC kinematics. (For related work on charmonium and bottomonium production we refer to \cite{Boer:2012bt,Schafer:2013wca}.)  Since the relevant analytical formulas already exist within the literature, in Sec.~\ref{s:phen} we focus on the phenomenology and refer the reader to the appropriate papers on the underlying theory.  These numerical results are summarized in Sec.~\ref{s:sum}, and there we highlight again how AFTER can offer unique insight into TSSAs in proton-proton collisions, which is a truly fundamental observable to test pQCD at higher twist.

%
%
\section{Pion, photon, and jet TSSAs at AFTER} \label{s:phen}

\begin{figure}[t]
 \begin{center}
  \fig{0.93}{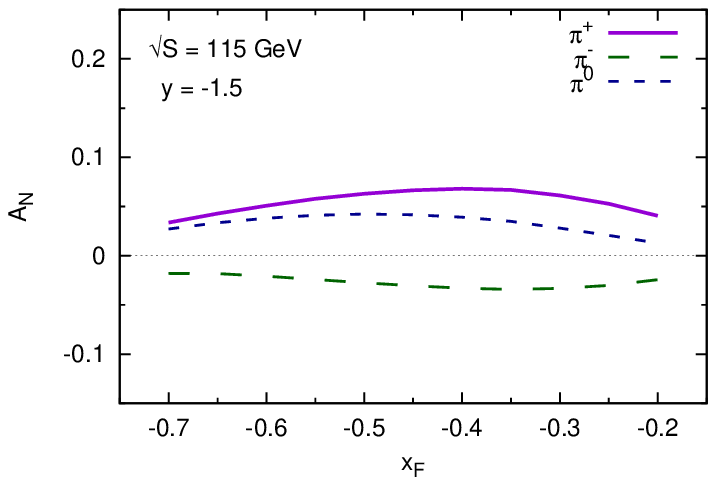}
  \fig{0.93}{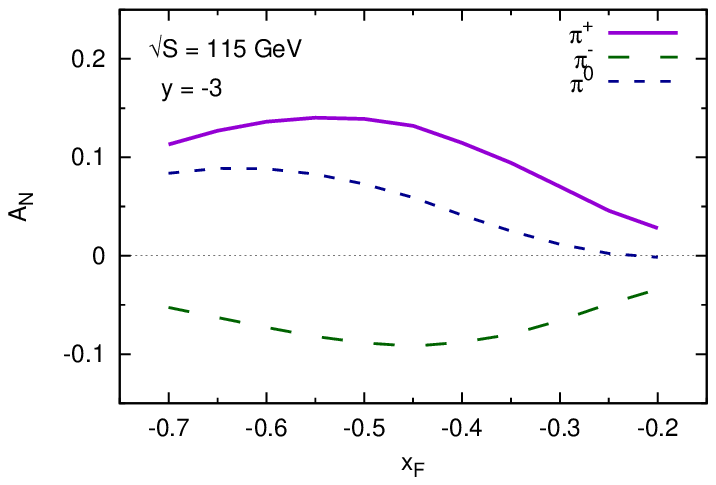}
  \fig{0.93}{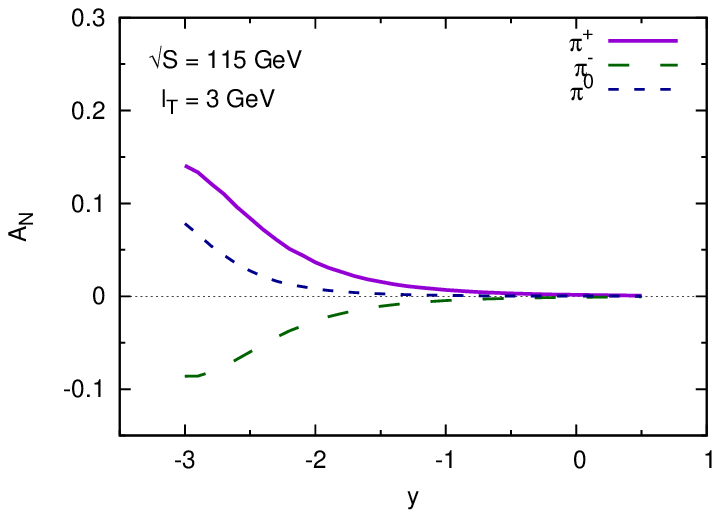}
 \end{center}
 \vspace{-0.9cm}
\caption{$A_N^\pi$ vs.~$x_F$ at fixed $y = -1.5$ (top left) and $y=-3$ (top right) as well as $A_N^\pi$ vs.~$y$ at fixed $l_T= 3\,{\rm GeV}$ (bottom).  All plots are at $\sqrt{S} = 115\,{\rm GeV}$ for pion production at AFTER.  
}
\label{f:AN_xF_pi}
\end{figure}

\begin{figure}[t]
 \begin{center}
  \fig{0.93}{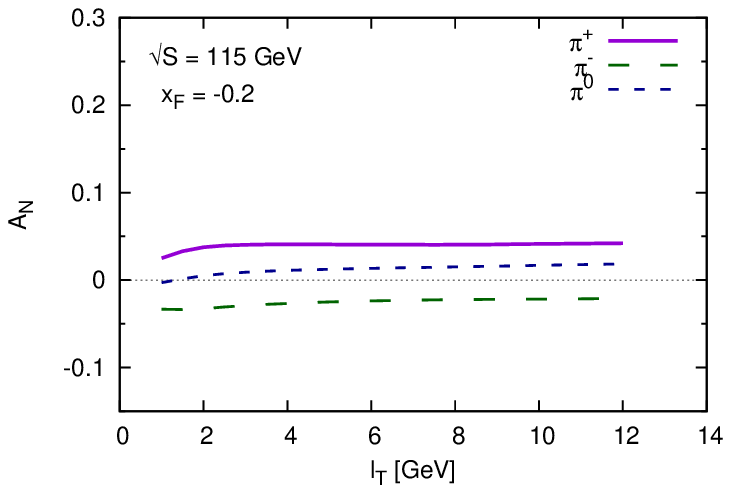}
  \fig{0.93}{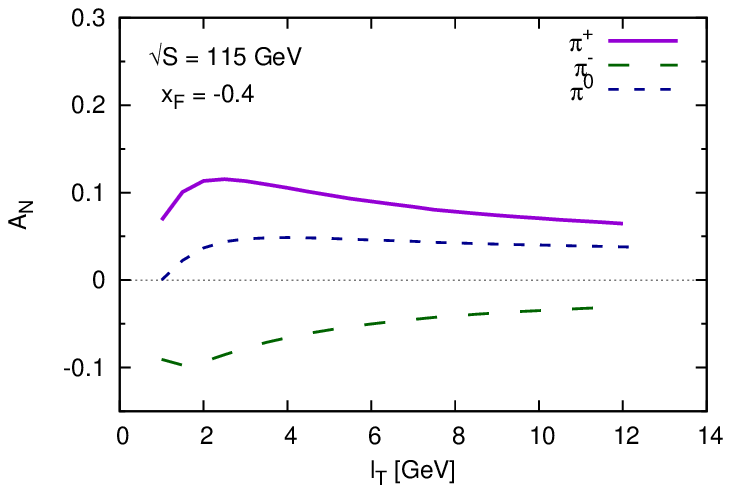}
  \fig{0.93}{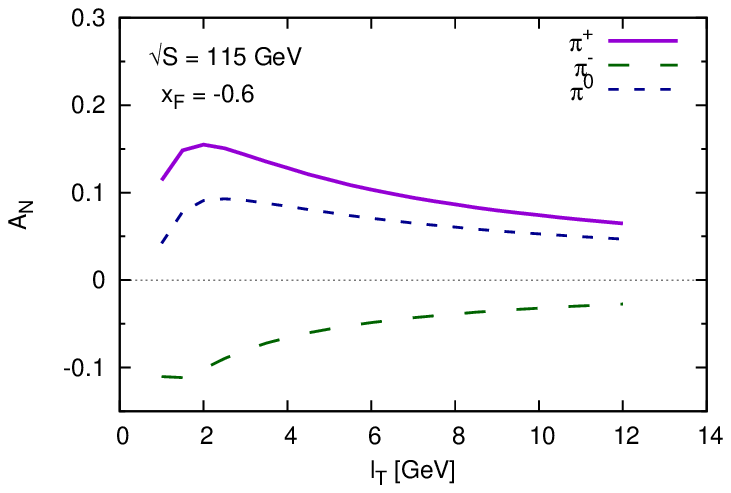}
 \end{center}
 \vspace{-1.1cm}
\caption{$A_N^\pi$ vs.~$l_T$ at fixed $x_F=-0.2$ (top left), $x_F = -0.4$ (top right), and $x_F= -0.6$ (bottom) at $\sqrt{S} = 115\,{\rm GeV}$ for pion production at AFTER.  
}
\label{f:AN_pT_pi}
\end{figure}

We start first with $A_N^\pi$, where we follow our numerical work in Ref.~\cite{Kanazawa:2014dca}.  (We also refer the reader to~\cite{Kouvaris:2006zy,Koike:2007rq,Metz:2012ct} for more formal discussions of the relevant analytical formulas.)  There we took into account the contribution from the QS function and the fragmentation term.  The former has a model-independent relation to the Sivers function~\cite{Boer:2003cm} while the latter involves three non-perturbative CT3 fragmentation functions (FFs):~$\hat{H}$, $\hat{H}^{\Im}_{FU}$, and $H$.  Of these, $\hat{H}$ has a model-independent connection to the Collins function~\cite{Kang:2010zzb,Metz:2012ct}, and $H$ can be written in terms of the other two through a QCD equation-of-motion relation~\cite{Metz:2012ct}.  In Figs.~\ref{f:AN_xF_pi},$\,$\ref{f:AN_pT_pi} we provide predictions for neutral and charged pion production at AFTER based on our fit in~\cite{Kanazawa:2014dca}.  One sees in Fig.~\ref{f:AN_xF_pi} from $A_N^\pi$ vs.~$x_F$ that the magnitude of the asymmetry can be anywhere from $\sim5-10\%$, and from $A_N^\pi$ vs.~$y$ that it increases with more negative (center-of-mass) rapidity $y$.\footnote{Recall the relation between $x_F$, $y$, and $l_T$: $x_F = 2\,l_T\sinh(y)/\sqrt{S}$, so $A_N$ vs.~$x_F$ ($A_N$ vs.~$y$) at fixed $y$ ($l_T$) implies a running in $l_T$ ($x_F$).}  One also notices that $A_N^\pi$ turns over at more negative $x_F$ values, which was also observed in some of the STAR data~\cite{Adams:2003fx,Adamczyk:2012xd}.  In Fig.~\ref{f:AN_pT_pi}, where we show $A_N^\pi$ vs.~$l_T$, one sees that the asymmetry is flat or falls off very slowly as $l_T$ increases, a feature that had also been measured by STAR~\cite{Heppelmann:2013ewa}.  It will be important to establish with more precision if this flatness persists at higher-$l_T$ values, say $12-15\,{\rm GeV}$, and AFTER, with its much higher luminosity, will be in a position to make such a measurement. 

\begin{figure}[H]
 \begin{center}
  \fig{0.92}{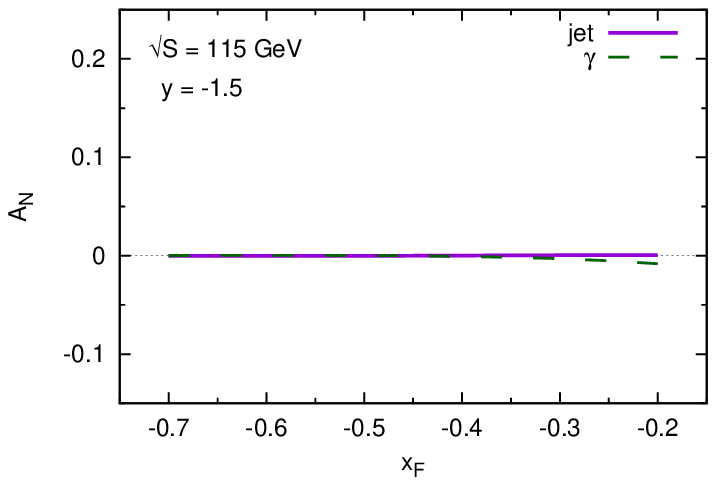}
  \fig{0.92}{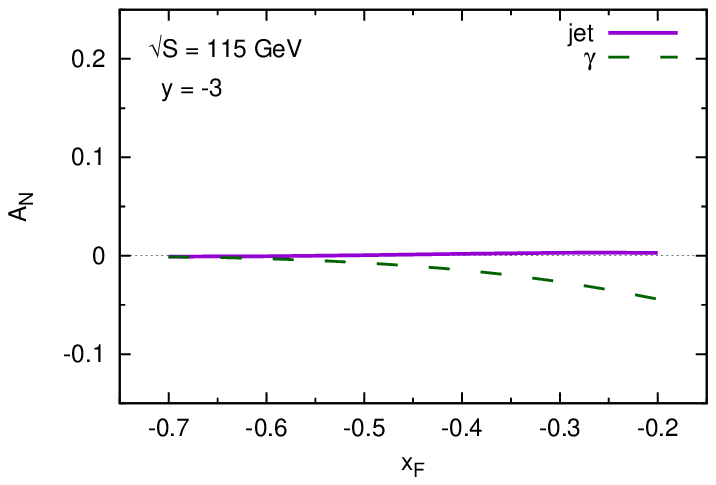}
  \fig{0.92}{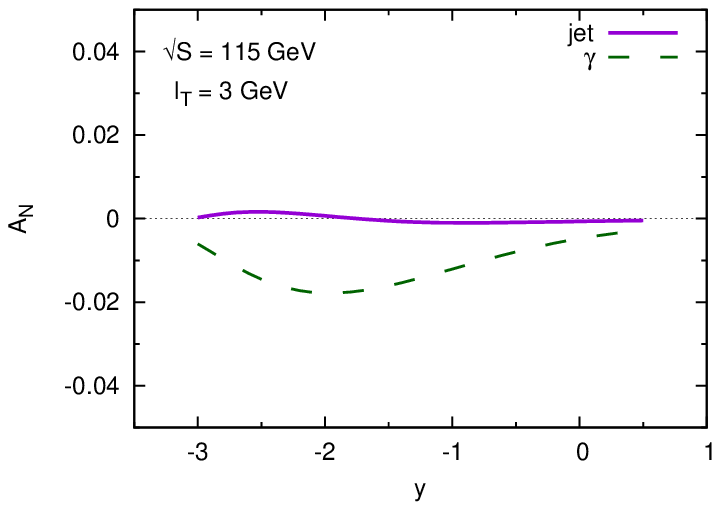}
 \end{center}
 \vspace{-0.9cm}
\caption{$A_N$ vs.~$x_F$ at fixed $y = -1.5$ (top left) and $y=-3$ (top right) as well as $A_N$ vs.~$y$ at fixed $l_T= 3\,{\rm GeV}$ (bottom).  All plots are at $\sqrt{S} = 115\,{\rm GeV}$ for jet/photon production at AFTER. \vspace{-0.5cm}
}
\label{f:AN_xF_jg}
\end{figure}

\begin{figure}[H]
 \begin{center}
  \fig{0.92}{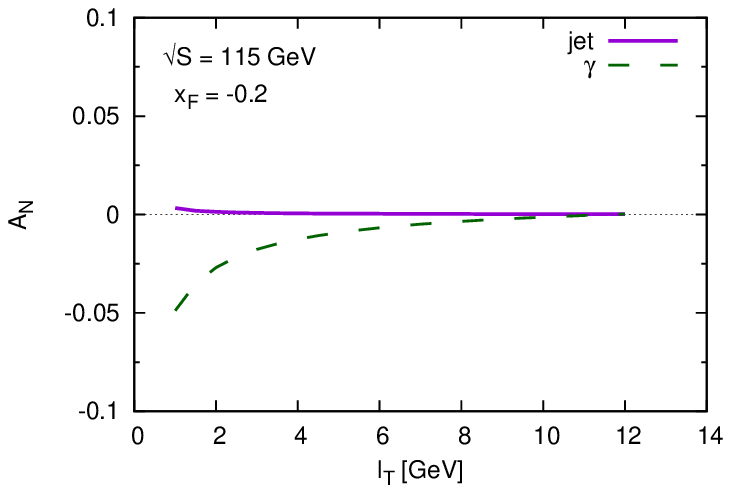}
  \fig{0.92}{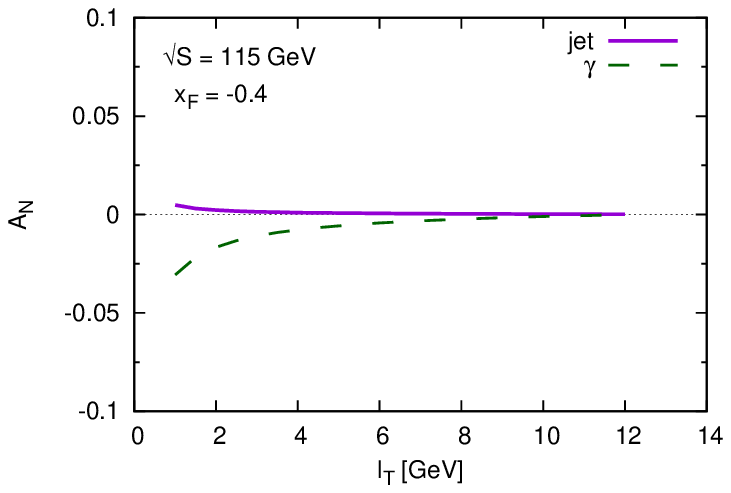}
  \fig{0.92}{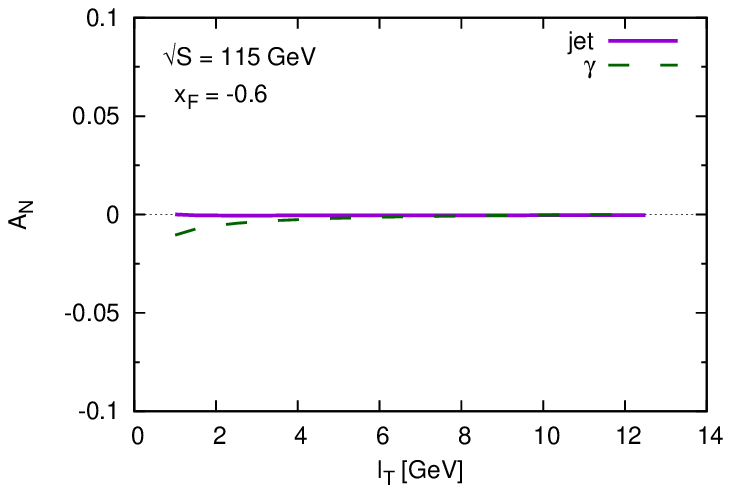}
 \end{center}
 \vspace{-1.1cm}
\caption{$A_N$ vs.~$l_T$ at fixed $x_F= -0.2$ (top left), $x_F = -0.4$ (top right), and $x_F= -0.6$ (bottom) at $\sqrt{S} = 115\,{\rm GeV}$ for jet/photon production at AFTER.  
}
\label{f:AN_pT_jg}
\end{figure}

We next look at $A_N^{jet}$ and $A_N^\gamma$, which do not receive contributions from FFs.  As we discussed in Sec.~\ref{s:intro}, the former may receive non-negligible contributions from terms other than the QS function, while for the latter we recently showed in~\cite{Kanazawa:2014nea} that the QS function is the dominant piece to that asymmetry.  (All of the analytical expressions for $A_N^\gamma$ can be found in Ref.~\cite{Kanazawa:2014nea}\footnote{We note the analytical formulas for the piece involving chiral-odd functions is new from that work, while those involving chiral-even functions were derived before in the literature, and the relevant references are cited therein.}, while those for $A_N^{jet}$ are determined simply by setting $D_1(z)$ (the unpolarized FF) to $\delta(1-z)$ in the equations for $A_N^\pi$ given in~\cite{Kouvaris:2006zy,Koike:2009ge,Beppu:2013uda,Kanazawa:2000hz}.)  However, given that the other pieces for $A_N^{jet}$ are not reliably known, for that asymmetry we will only look at the contribution from the QS function using its relation to the Sivers function, while for $A_N^\gamma$ we adopt our work in~\cite{Kanazawa:2014nea}.  In Figs.~\ref{f:AN_xF_jg}, \ref{f:AN_pT_jg} we show results for jet and photon production at AFTER.  We see that $A_N^{jet}$ is very small, although we caution the reader that the Sivers function (which we use as input for the QS function) is mostly unconstrained in the large-$x^\uparrow$ region, and when this uncertainty is taken into account, one could obtain a measurable asymmetry~\cite{Gamberg:2013kla}.  Also, as we mentioned, there is the potential for (chiral-even) SFP and/or tri-gluon functions to make an impact.  Therefore, in order to determine if the Sivers function alone can describe $A_N^{jet}$, along with the current data from A$_N$DY, we need more precise data in the far backward region, which should be possible at AFTER.\footnote{We note that STAR has preliminary data on electromagnetic ``jets'' that could also be helpful~\cite{Mondal:2014vla}.}
 
Unlike the jet asymmetry, $A_N^\gamma$ could be on the order of $\sim\!-5\%$ at less negative $x_F$ and more negative $y$ (see Fig.~\ref{f:AN_xF_jg} (top)) or smaller $l_T$ and less negative $x_F$ (see Fig.~\ref{f:AN_pT_jg}).  Both of these observations are consistent with the behavior of $A_N^\gamma$ as a function of rapidity (see Fig.~\ref{f:AN_xF_jg} (bottom)), where the asymmetry peaks at $y\sim -2$ (with $l_T = 3\,{\rm GeV}$), which corresponds to $x_F\sim -0.2$.  Since the QS function is the dominant source of the asymmetry, we can have ``clean'' access to it.  We state again that the GPM framework at present predicts $A_N^\gamma$ to be {\it positive}~\cite{Anselmino:2013rya}.  Therefore, a clear nonzero signal for this observable would help to distinguish between the CT3 and GPM formalisms.  However, we emphasize that should data contradict the predictions of the GPM, this does not invalidate the results obtained for TMD observables that are based on rigorous TMD factorization proofs. Also, since we use the Sivers function from SIDIS as our input for the QS function, we can learn about the predicted process dependence of the Sivers function.  

%
%
\section{Summary and outlook} \label{s:sum}

In this paper we have discussed TSSAs in single-inclusive pion, jet, and photon production from proton-proton collisions, i.e., $p^\uparrow p\to \{\pi,\,jet,\,\gamma\} \,X$, at kinematics relevant for the proposed AFTER@LHC experiment.  These asymmetries have been fundamental observables to test pQCD at higher twist for close to 40 years, and much work has been performed on both the theoretical and experimental sides.  Nevertheless, issues still remain as to the origin of these TSSAs, which makes a measurement of $A_N$ at the LHC via the AFTER experiment timely. For $A_N^\pi$ we have found that AFTER should expect (absolute) asymmetries on the order of $~5-10\%$ as a function of $x_F$ and increasing as the rapidity becomes more negative.  Also, the $l_T$ dependence of $A_N^\pi$ still falls off slowly and flattens out at high $l_T$.  For $A_N^{jet}$ we predict a very small asymmetry, but we must remember that uncertainties in the Sivers function could allow for a measurable observable~\cite{Gamberg:2013kla} and also that other contributions (like chiral-even SFP and tri-gluon) could make an impact.   Lastly, for $A_N^\gamma$ we expect asymmetries on the order of $\sim-5\%$ and decreasing with more negative $x_F$ and increasing $l_T$.  These are opposite in sign to the ones predicted from the GPM~\cite{Anselmino:2013rya}. 

Even though these observables have been (or are planned to be) measured at RHIC, AFTER has the ability, through its much higher luminosity, to not just supplement the RHIC data, but also provide important information on still unknown issues.  For example, it will be key to determine if $A_N^\pi$ stays flat at higher-$l_T$, say to 12-15 GeV, like theory predicts~\cite{Kanazawa:2012kt,Kanazawa:2014dca,Anselmino:2013rya,Anselmino:2015eoa,Anselmino:2000vs} and STAR has evidence for~\cite{Heppelmann:2013ewa}.  Also, higher statistics should allow for more precise measurements of $A_N^{jet}$ at more negative $x_F$, which will be necessary to determine if the QS function is the sole source of that asymmetry. Moreover, $A_N^\gamma$ has never been measured before and provides the opportunity to clearly distinguish between the CT3 and GPM frameworks and learn about the process dependence of the Sivers function.  Given the questions that remain as to the origin of TSSAs, which has been unresolved for almost 40 years, AFTER could provide valuable data on these observables.


%
%
\section*{Acknowledgments}

This work has been supported by the Grant-in-Aid for
Scientific Research from the Japanese Society of Promotion of Science
under Contract No.~26287040 (Y.K.), the National Science
Foundation under Contract No.~PHY-1205942 (K.K. and A.M.), and the RIKEN BNL
Research Center (D.P.).

\end{document}